\documentclass[pra]{revtex4}

\usepackage{graphicx}

\begin{document}


\newcommand{\nc}{\newcommand}
\newcommand{\BE}{\begin{equation}}
\newcommand{\EE}{\end{equation}}
\newcommand{\BA}{\begin{eqnarray}}
\newcommand{\EA}{\end{eqnarray}}

\title{ Tunneling Time Distribution by means of Nelson's Quantum Mechanics and Wave-Particle Duality }

\author{
Koh'ichiro Hara$^{1}$\footnote{E-mail khara@hep.phys.waseda.ac.jp},
and Ichiro Ohba$^{1,2,3}$\footnote{E-mail ohba@mn.waseda.ac.jp}
}

\address{
$^1$Department of Physics, Waseda University Tokyo 169-8555, Japan,\\
$^2$Kagami Memorial Laboratory for Materials Science and Technology,\\
Waseda University Tokyo 169-0051, Japan,\\
$^3$Advanced Research Center for Science and Technology,\\
Waseda University, Tokyo 169-8555, Japan
}

\begin{abstract}
We calculate a tunneling time distribution by means of Nelson's 
quantum mechanics and investigate its statistical properties.
The relationship between the average and deviation of tunneling time
suggests the existence of ``wave-particle duality'' in the tunneling 
phenomena.
\end{abstract}


\pacs{03.65.Xp, 73.40.Gk}

\maketitle

   \section{Introduction}

It was suggested that there is a time associated with the passage of 
a particle under a tunneling barrier, so-called tunneling time
\cite{MacColl}.
Actually, several authors have tried to measure the time experimentally
\cite{Martinis,Gueret=1988a,Gueret=1988b}. However, 
there is no clear consensus about any definition of tunneling time
admissible for everyone.
There are several approaches and methods to estimate
the tunneling time. For example, Wigner time 
\cite{Hartman,Hauge=1987,Jaworski} is based on 
the time evolution of wave packet through the barrier,
and the delay time of the peak or the centroid is expressed by an
energy derivative of phase shift. Larmor time 
\cite{Baz=1967a,Baz=1967b,Rybachenko,Verhaar,Buttiker=1983}
 is obtained from the Larmor precession angle caused by a magnetic 
field confined in the barrier region.
The traversal time proposed by B\"uttiker and 
Landauer \cite{Buttiker=1982,Buttiker=1985,Martin=1993} is defined 
by the analysis of transmission coefficient through a static barrier 
augmented by a small oscillation in the barrier height. 
This time is obtained by measuring or analyzing
the effect of ``clock'' added on the tunneling barrier.
The dwell time \cite{Smith,Jauch,Buttiker=1983} is defined as 
the total probability of the particle within the barrier 
divided by the incident probability current.
This is only applicable to the stationary state case.
There are another type of methods based on the motion of ``particle paths'',
for example, Bohmian mechanics 
\cite{Leavens=1993a,Leavens=1993b,McKinnon}, 
Feynman path integral \cite{Sokolovski,Fertig,Yamada=2000}, Nelson's 
quantum mechanics \cite{Imafuku=1995,Imafuku=1997,Hara=2000}, 
and so on. In these methods, the tunneling 
time is defined as the time spent by ``particle paths'' under 
the tunneling barrier.
See Refs. \cite{Hauge=1989,Landauer=1994} and references therein
for reviews of this problem.

We think that Nelson's quantum mechanics \cite{Nelson=1966}
have some characteristic properties to study the tunneling time 
as follows. Since this method is described by the real-time stochastic process,
it enables us to describe the real-time evolution of individual events as the
analogy of classical mechanics. We call such an event of evolution as
a ``sample path.''
Moreover, since a sample path has its own history, we can obtain 
information of time parameter, in particular, tunneling time.

In this paper, first, we calculate a tunneling time distribution 
by means of Nelson's quantum mechanics \cite{Nelson=1966}.
The tunneling phenomena should occur quantum mechanically 
and accompany fluctuating properties.
If a tunneling particle is described by a wave packet with distribution of 
wavenumbers of finite width, then one would have a distribution of 
tunneling time. The width of the distribution should be reduced by using
the spatially wider wave packet or the sharper distribution of 
wavenumber wave packet.
In such case there should still remain the fluctuation of tunneling time
coming from quantum effect.
In spite of such an argument, many approaches centered discussion 
only on the averaged value, because it is difficult to consider 
the tunneling time distribution by the use of the conventional frameworks 
of quantum mechanics. However, Nelson's quantum mechanics can afford
to predict such distribution, because this method enables us to 
obtain an ensemble of various sample paths in tunneling barrier.
Next, from the tunneling time distribution calculated by 
this method, we investigate the statistical properties of the 
distribution, such as the average and the deviation
of tunneling time. We found that the relationship between
the deviation and average suggests the existence of the 
 ``wave-particle duality'' in the tunneling phenomena.
Last, we discuss the ``quantum-classical correspondence'' by 
analyzing the ``Planck constant dependence'' of tunneling time distribution.

In this paper, for simplicity, we analyze the tunneling phenomena 
with one-dimensional static rectangular potential barrier with height 
$V_0$ and width $d$ as following,
\begin{eqnarray}
V(x) & = & \left\{
\begin{array}{lll}
0 & x<-d/2 & \mbox{(region I)},  \\
V_0 & -d/2 \leq x \leq d/2 & \mbox{(region II)},  \\
0 & x>d/2 & \mbox{(region III)}.  \\
\end{array}
\right.
\end{eqnarray}
The initial Gaussian wave packet with variance $\Delta x^2$,
\begin{equation}
\psi(x,0) = \left( \frac{1}{2 \pi (\Delta x)^2} \right)^{1/4} 
\exp \left[-\frac{\left(x-\langle x\rangle \right)^2}{4(\Delta x)^2}
+\frac{i}{\hbar} \langle p \rangle \left(x- \langle x\rangle \right)
\right],
\label{initial}
\end{equation}
is injected into the potential barrier, where $\langle x \rangle$, and $\langle p \rangle$ are an 
expectation value of position and momentum respectively.
We use the natural unit $m=\hbar=1$, and perform numerical 
simulations of 100,000 sample paths.
For the numerical simulation, we take $[-1000,1000]$ as the total space 
for the unit of $1/k_0$, and adopt these parameters as follows,
$\langle x \rangle=-500/k_0$, 
and $\langle p \rangle=k_0$.

\section{Estimation of the tunneling time based on the Nelson's quantum mechanics}

In this section, we give a brief review of the Nelson's approach
of quantum mechanics \cite{Nelson=1966} which plays a central role in this paper, 
and explain how to estimate the tunneling time.
Nelson's quantum mechanics based on the real-time stochastic process,
enables us to describe the quantum mechanics of a single particle 
in terminology of the ``analog'' of classical mechanics,
i.e. the ensemble of sample paths. These sample paths are generated
by the Ito type Langevin equation,
\begin{equation}
dx(t) = [ u (x(t),t) + v (x(t),t) ]dt + dw(t),
\label{Langevin}
\end{equation}
where $x(t)$ is a stochastic variable corresponding to the 
coordinate of the particle, and $u (x(t),t)$ and $v(x(t),t)$ are 
the osmotic velocity and the current velocity, respectively.
The $dw(t)$ is the Gaussian white noise with the statistical properties of
\begin{equation}
\langle dw(t) \rangle = 0, \ \ \mbox{and} \ \ 
\langle dw(t) dw(t) \rangle = \frac{\hbar}{m} dt,
\end{equation}
where $\langle \cdots \rangle$ means the ensemble average with
respect to the noise. In principle, the osmotic and the current 
velocities are given by solving coupled two equation, i.e. the 
kinetic equation and the ``Nelson--Newton equation''. Nelson showed that,
for the expectation value of the dynamical variable, e.g., $x, p$, 
the whole ensemble of sample paths gives us the same results 
as quantum mechanics in the ordinary approach \cite{Nelson=1966}.
Once the equivalence between Nelson's framework and the ordinary 
quantum mechanics is proved, it is convenient to use the relation
\begin{equation}
u={\rm Re} \frac{\hbar}{m} \frac{\partial}{\partial x} 
\ln \psi (x,t) , \ \ \mbox{and} \ \ 
v={\rm Im} \frac{\hbar}{m} \frac{\partial}{\partial x} 
\ln \psi (x,t),
\label{verocity}
\end{equation}
where $\psi$ is the solution of Schr\"odinger equation. 
Since an individual sample path has its own history, we obtain 
information on the time parameter, e.g. tunneling time definitely.

Now using the Nelson's quantum mechanics, we estimate the 
tunneling time of a particle crossing over a potential barrier.
First, we prepare an incident wave packet given by Eq.(\ref{initial}) 
from the region I. Next, we solve the time-dependent 
Schr\"odinger equation.
Last, using the relation Eq.(\ref{verocity}), we obtain the drift term
of the Langevin equation (\ref{Langevin}), and calculate sample paths. 
Suppose a simulation of tunneling phenomena based on Eq.(\ref{Langevin}),
starting for $t=-\infty$ and ending $t=\infty$. As we treat 
a wave packet satisfying the time-dependent Schr\"odinger 
equation, the wave packet is located region I 
initially and turns finally into two spatially separated wave
packets which are regions I and III, respectively. Figure 1 
shows a typical transmitted sample path calculated by Eq.(\ref{Langevin}).
Transmitted sample paths originate preferentially from 
the front of the initial wave packet as suggested 
by Imafuku et al.\cite{Imafuku=1995}.
Every transmitted sample path has its traversal time of barrier,
i.e., the tunneling time which is described as 
\BE
\tau_{i} \equiv \int^{t_f}_0 \Theta(x_i(t)) dt, \ (i=1, 2, \cdots, N),
\EE
where $x_i(t)$ is the $i$-th sample path, $t_f$ is the final time,
the function $\Theta(x)$ is unity for $-d/2 \le x \le d/2$ and zero otherwise.
Collecting these events, we can 
construct a statistical distribution of tunneling time.

\section{Tunneling time distribution and wave-particle duality}

From the ensemble of sample paths, we define a distribution of tunneling
time as follows,
\begin{equation}
P(\tau) \delta \tau \equiv \frac{\delta n(\tau)}{N},
\end{equation}
$\delta n(\tau)$ is the number of sample paths with the tunneling 
time from $\tau$ to $\tau + \delta \tau$, and $N$ is the total number
of sample paths.

Figures 3 and 4 show the average of tunneling time $\langle \tau \rangle$
 and its deviation $\Delta \tau$ versus potential barrier width for 
various width of wave packet $\Delta x$.
We can see that in the case of $\Delta x$ greater than $20/k_0$, 
behaviors of average and deviation are independent on $\Delta x$.
However, in the case of $\Delta x = 10/k_0$, the data deviate remarkably 
from the others in the region $d \ge 10/k_0$. It is suggested that this effect
originates from the wave packet spreading during the propagation.
Indeed the spatial deviation of free wave packet at time $t$ is 
$\sqrt{\Delta x^2 + \frac{\hbar^2}{4 m^2 \Delta x^2}t^2}$.
Therefore, the wave packet spreads twofold at $t=2m(\Delta x)^2/\hbar$.
By taking account of the fixed time $t=500/k_0^2$ in which the peak of
initial wave packet arrives at the left edge of barrier,
this occurs in the case of width
$\Delta x \le \sqrt{\frac{250}{m k_0^2}\hbar} \sim 16/k_0$.

It seems that the numerical results of $\langle \tau \rangle$ are roughly
similar to the WKB time $\tau_{\rm WKB}=md/\hbar \kappa$, where $\kappa=\sqrt{2m(V_0-E_0)/\hbar^2}$, expect for the case of $\Delta x=10/k_0$.
However, let us examine much in detail those values in the thin 
region ($d \le 4/k_0$). Figure 5 is an enlarged copy of this part in Fig.3.
It has been shown that, in the opaque case, the numerical simulation gives almost same values of the WKB times \cite{Imafuku=1995,Imafuku=1997}.
While we can see these features of opaque case in Figs. 3 and 5,
the numerical values deviate from the WKB time in the translucent case 
characterised by small $\kappa d$ which is approximately less than $2$. 
The WKB approximation is not proper in the latter case.
This suggests the tunneling phenomena make a ``phase transition''
in a sense around $\kappa d \sim 2$.

Next we show in Fig.6 the deviation $\Delta \tau$ as a function of
the average $\langle \tau \rangle$, which is calculated by changing
the width $d$ with fixed potential height $V_0$ for several cases of 
incident energy $E_0$. For each case we can see 
a common feature characterized by the fact that $\Delta \tau$ is 
proportional to $\langle \tau \rangle$ for 
$\kappa d \le 2$ and to $\sqrt{\langle \tau \rangle}$
for $\kappa d \ge 2$. In order to check
this feature quantitatively, we fit the tunneling time distribution 
using the Gamma distribution,
\begin{equation}
P(\tau)=\frac{1}{\beta^{\alpha + 1} \Gamma (\alpha +1)}\tau^{\alpha}
e^{-\tau/\beta} \ \ (\alpha, \beta, \tau >0).
\end{equation}
This distribution has the following statistical properties,
\begin{equation}
\langle \tau \rangle=\beta(\alpha + 1), \ \mbox{and} \
\Delta \tau^2= \beta^2 (\alpha + 1).
\end{equation}
If $\alpha$ is constant, the relation 
$\Delta \tau \propto \langle \tau \rangle$ holds good, and this is 
a typical feature of such coherent phenomena as deviation versus 
average value of photon number in coherent photon state.
On the other hand, if $\beta$ is constant, the relation $\Delta \tau 
\propto \sqrt{\langle \tau \rangle}$ holds good, and this is a typical
feature of such random phenomena as Poisson process and Brownian motion.
Figure 7 is an example of fitting distribution.

Figures 8 and 9 show the fitting parameters $\alpha$ and $\kappa^2 \beta$
as a function of $\kappa d$, respectively.
We see the $\kappa d$ dependence of fitting parameters
is universal even for different potential heights.
Note that $\kappa$ is kept a constant value for each case.
Furthermore, we found that $\alpha$ is constant in the translucent region
($\kappa d \le 2$), and $\kappa^2 \beta$ is constant
in the opaque region ($\kappa d \ge 2$).
Therefore, this assures the $\Delta \tau$-$\langle \tau \rangle$ relation
as mentioned above.

Anybody never doubts that the tunneling time is a pure quantum process.
However, the tunneling time is not an observable in the quantum mechanics.
Actually it can be closely connected to the time evolution of some specific
observable of the tunneling system and it should be measured experimentally
through the time dependence of the observable.
Thus the statistical property of the tunneling time distribution should
reflect the characteristic features of underlying quantum process.
From the above discussion
we think that the tunneling may occur coherently, or in mode of
wave picture dominantly in the translucent case and randomly, or
in mode of particle picture dominantly in the opaque case.
The tunneling time distribution reveals the wave-particle duality
in the tunneling phenomena.

\section{Quantum-classical correspondence}

We discuss the Planck constant dependence of tunneling time
distribution. First, we introduce the parameter 
$\epsilon \ (0< \epsilon \leq 1)$ and the ``Planck constant''
$\tilde{\hbar}=\epsilon \hbar$, and consider ``Schr\"odinger equation''
of a wave function $\tilde{\psi}(x,t)$,
\begin{equation}
i\tilde{\hbar}\frac{\partial}{\partial t} \tilde{\psi}(x,t)=
\left[ -\frac{\tilde{\hbar}^2}{2m}\frac{\partial^2}{\partial x^2} +V(x) \right]
\tilde{\psi}(x,t),
\label{tSE}
\end{equation}
under the conditions with fixed values of 
$m, V_0, \mbox{and} \ p_0(=\hbar k_0)$.
This wave function describes a virtual quantum system with scaled Planck
constant $\tilde{\hbar}$.
This equation is formally transformed into
 the ordinary Schr\"odinger equation
\begin{equation}
i\hbar\frac{\partial}{\partial T} \Phi(X,T)=
\left[ -\frac{\hbar^2}{2m}\frac{\partial^2}{\partial X^2} +W(X) \right]
\Phi(X,T),
\label{tSEII}
\end{equation}
by a scale transformation as following:
\begin{equation}
X \equiv x/\epsilon, \ T \equiv t/\epsilon, \ W(X) \equiv V(\epsilon X),
\mbox{and} \ \Phi(X,T) \equiv \tilde{\psi}(\epsilon X, \epsilon T).
\end{equation}
Here let us clear up the procedure of the simulation.
We fix the parameters $m, V_0, E_0=p^2/2m,$ and $d$.
We scale the space-time variables as $X=x/ \epsilon,$ $T=t/ \epsilon$ and
the potential width as $D=d/ \epsilon$. In this $X-T$ reference frame
de Broglie plane wave is scaled as
$e^{i(kx-\omega t)} \rightarrow e^{i(\tilde{k} X- \tilde{\omega} T)},$
where $\tilde{k}=\epsilon k$ and $\tilde{\omega}=\epsilon \omega$.
Then we perform the numerical simulation based on Eq.(\ref{tSEII})
and obtain the distribution of $\tau$ in the $X-T$ reference frame.
Eventually we obtain the distribution of $\tilde{\tau},$ the tunneling time
corresponding to a Planck constant $\tilde{\hbar}$, by the scaling of
$\tilde{\tau}=\epsilon \tau$.

Now we will examine how the average 
$\langle \tilde{\tau} \rangle=\epsilon \langle \tau \rangle$ and 
the deviation $\Delta \tilde{\tau}=\epsilon \Delta \tau$ depend on $\hbar$.
Let us start the simulation at a translucent case with $\epsilon=1$,
where the tunneling process proceeds in the wave mode.
Then, we decrease the value of $\epsilon$ gradually, the effective
width of tunneling barrier becomes wider and the tunneling process
should proceed in the particle mode, or in other words,
quasi-classically.
This procedure gives us the $\epsilon$-dependence of 
$\langle \tilde{\tau} \rangle$ shown in Fig.10.
We can see that $\langle \tilde{\tau} \rangle$ increases with a tendency
to approach the WKB time as $\epsilon$ is decreased in the region near
$\epsilon =1$, whereas it is almost independent of $\epsilon$ in the region
with smaller $\epsilon$. These tendencies may be understood as follows.
As it is seen in Fig.5, the simulated values of $\langle \tau \rangle$ fit
almost to the WKB time,
\begin{equation}
\tau_{\rm WKB}=\sqrt{\frac{m}{2(V_0-E_0)}}D,
\end{equation}
in the latter case. Thus, the tunneling time corresponding to $\tilde{\hbar}$
is $\langle \tilde{\tau} \rangle=
\epsilon \tau_{\rm WKB}=\sqrt{\frac{m}{2(V_0-E_0)}}d$.
Actually note that the WKB time can be expressed only by classical
quantities.
On the other hand, in the translucent case, the simulated values are 
smaller than the WKB time and approach gradually it 
as the effective width of tunneling barrier becomes wider,
that is, $\epsilon$ is decreased. Moreover, these two tendencies cross
each other at $\epsilon \sim 1/2$ which suggests 
the tunneling phenomena change their phase around 
$\hbar \kappa d \sim 2 \tilde{\hbar}$.

Next, let us consider the $\epsilon$-dependence of $\Delta \tilde{\tau}$.
In this simulation, we may guess that in the region near $\epsilon=1$
(wave mode and $\Delta \tau \propto \langle \tau \rangle$) the deviation
should be proportional to the average,
\begin{equation}
\Delta \tilde{\tau} \propto \langle \tilde{\tau} \rangle,
\end{equation}
whereas in the region with smaller $\epsilon$ (particle mode and
$\Delta \tau \propto \sqrt{\langle \tau \rangle}$), it should behaves as
\begin{equation}
\Delta \tilde{\tau} \propto 
\sqrt{\epsilon} \sqrt{\langle \tilde{\tau} \rangle}.
\end{equation}
Moreover, the change from the former tendency to the latter one may occur
around $\hbar \kappa d \sim 2 \tilde{\hbar}$.

We can see these tendencies in Fig.11, which give another support of the
idea of ``wave-particle duality'' in the tunneling phenomena.

\section{Summary}

We calculate the tunneling time distribution by means of 
Nelson's quantum mechanics. From the resulting distribution,
we derived the statistical properties of it,
the average and deviation of tunneling time.
First, we found that if an incident wave packet is so large as to look like plane-wave like, the $\Delta x$ dependence of them is negligible.
Next, fitting the data by Gamma distribution, we found that
the shape of distribution is universally determined only by 
$\kappa d$.
Furthermore, by investigating the statistical properties of the
distribution in two characteristic ``translucent''
and ``opaque'' regions roughly divided by $\kappa d \sim 2$,
we found that the ``wave-particle duality'' may be seen in the tunneling 
phenomena.
Last, we consider the Planck constant dependence of tunneling time,
introducing the parameter $\epsilon$ as $\tilde{\hbar}=\epsilon \hbar$.
Consequently, we found that the dependences of the average and the 
deviation suggest another support of the idea of ``wave-particle duality" 
in tunneling phenomena.

We are interested in the comparison between the tunneling time distribution
based on the Nelson's quantum mechanics and that based on the other method,
e.g., Bohmian mechanics. The detailed study of such a comparison is a subject
in the near future.

\section{Acknowledgements}

This work is supported by the Grant-in-Aid for COE Research
and that for Priority Area B (\#763), MEXT.


\newpage

\begin{figure}
\includegraphics{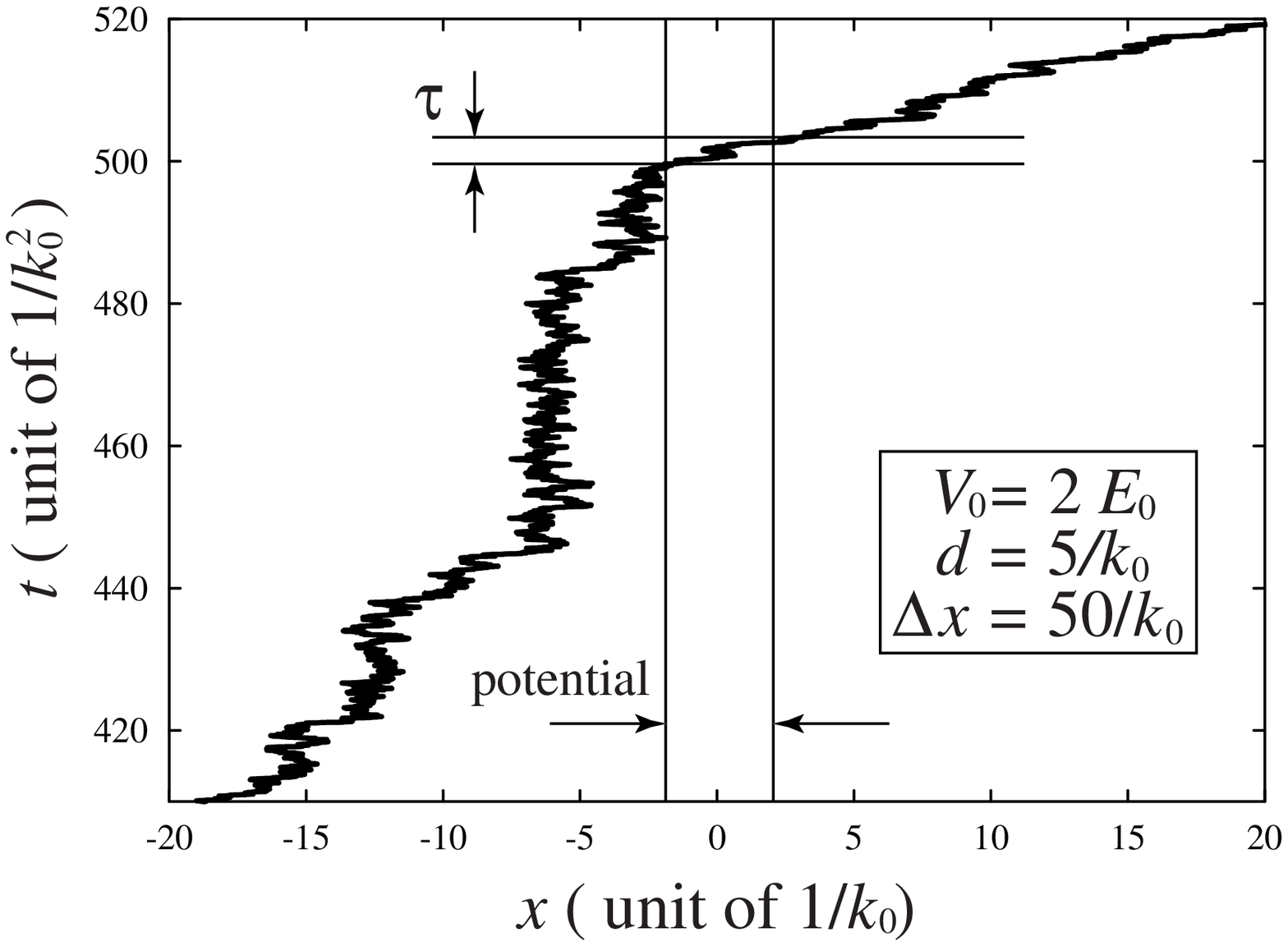}
\caption{\label{fig1} Typical transmitted sample path calculated by Eq.(\ref{Langevin}).
$\tau$ shows the traversal time of barrier, i.e., the tunneling time.}
\end{figure}

\newpage

\begin{figure}
\includegraphics{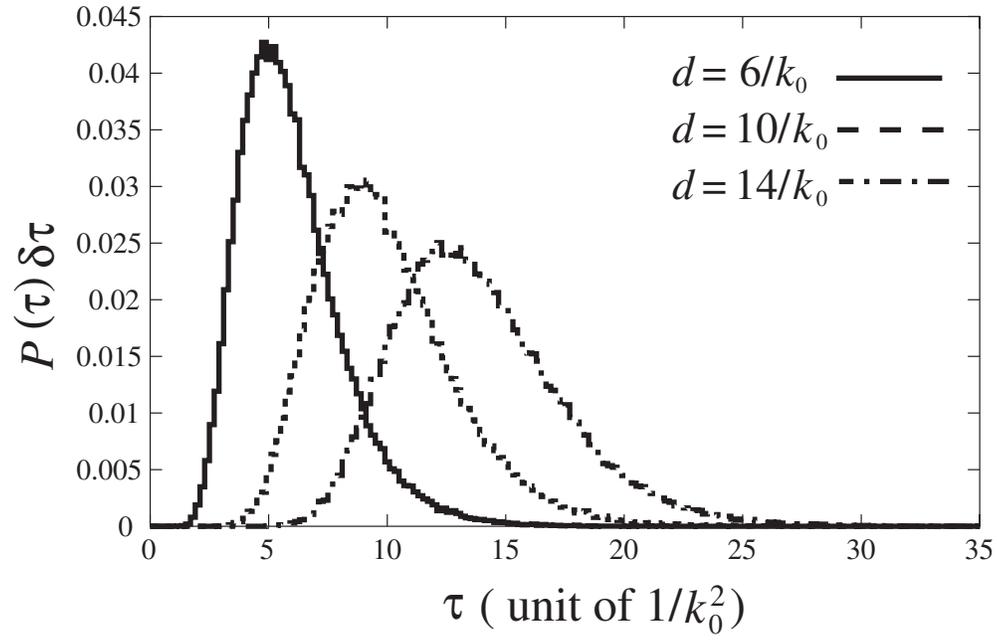}
\caption{\label{fig2} Numerical results of tunneling time distribution
as a function of tunneling time with potential height $V_0/E_0=2.0$.}
\end{figure}

\newpage

\begin{figure}
\includegraphics{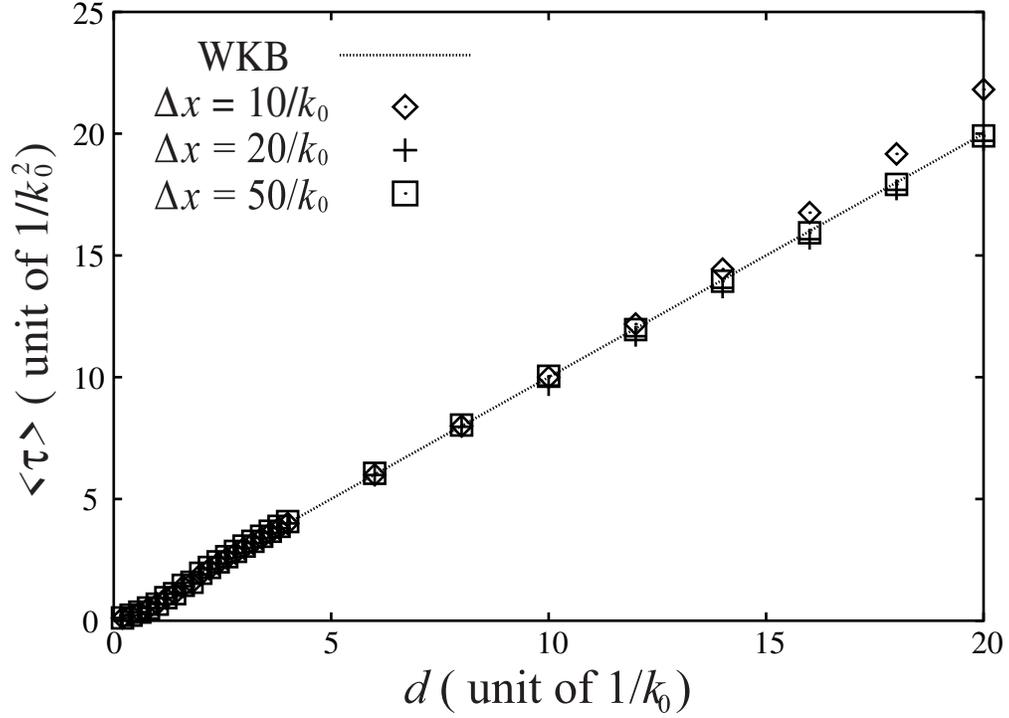}
\caption{\label{fig:3}Numerical results of tunneling time average versus 
potential width with potential height of $V_0/E_0=2.0$.}
\end{figure}

\newpage

\begin{figure}
\includegraphics{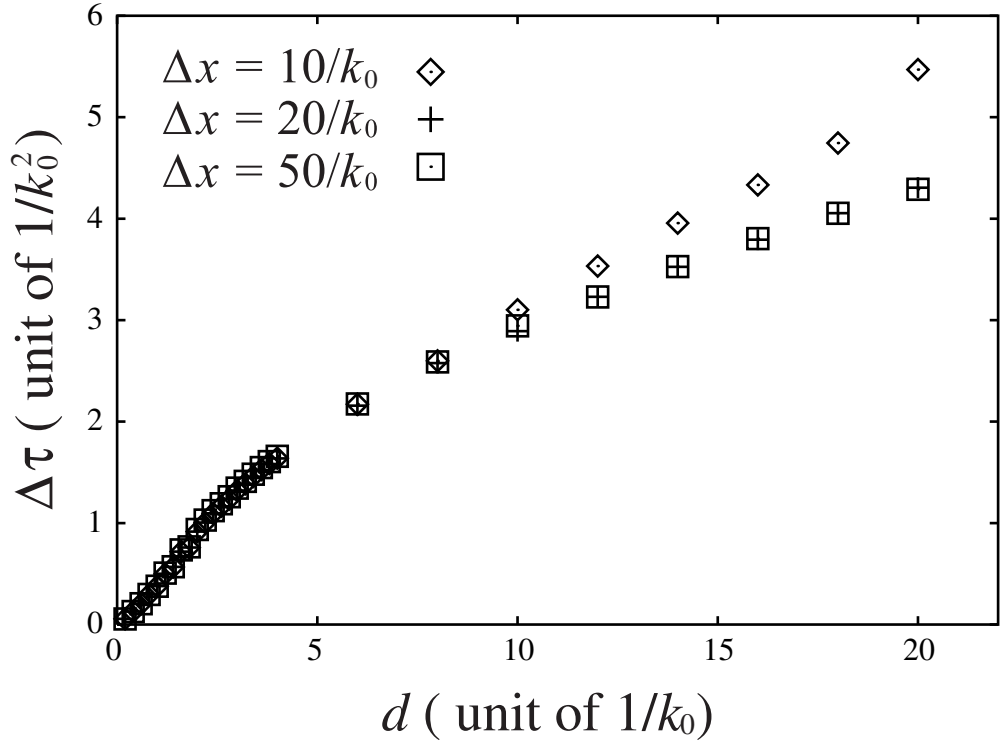}
\caption{\label{fig:4}Numerical results of tunneling time deviation versus 
potential width for potential height of $V_0/E_0=2.0$.}
\end{figure}

\newpage

\begin{figure}
\includegraphics{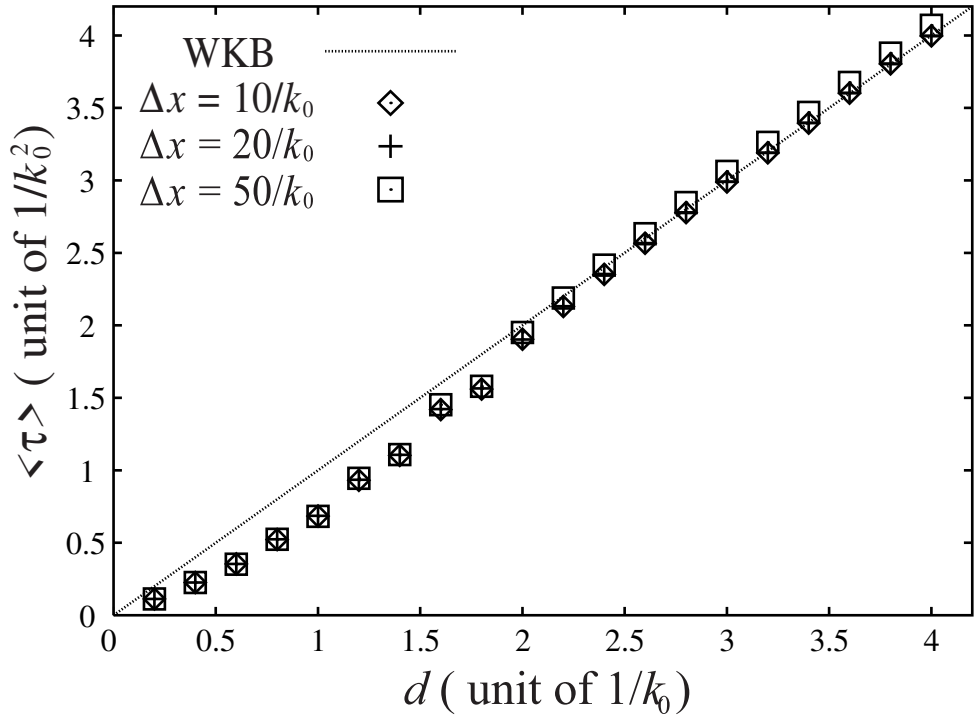}
\caption{\label{fig:5}An enlarged copy of thin barrier region ($d \le 4/k_0$) in Fig. 3.}
\end{figure}

\newpage

\begin{figure}
\includegraphics{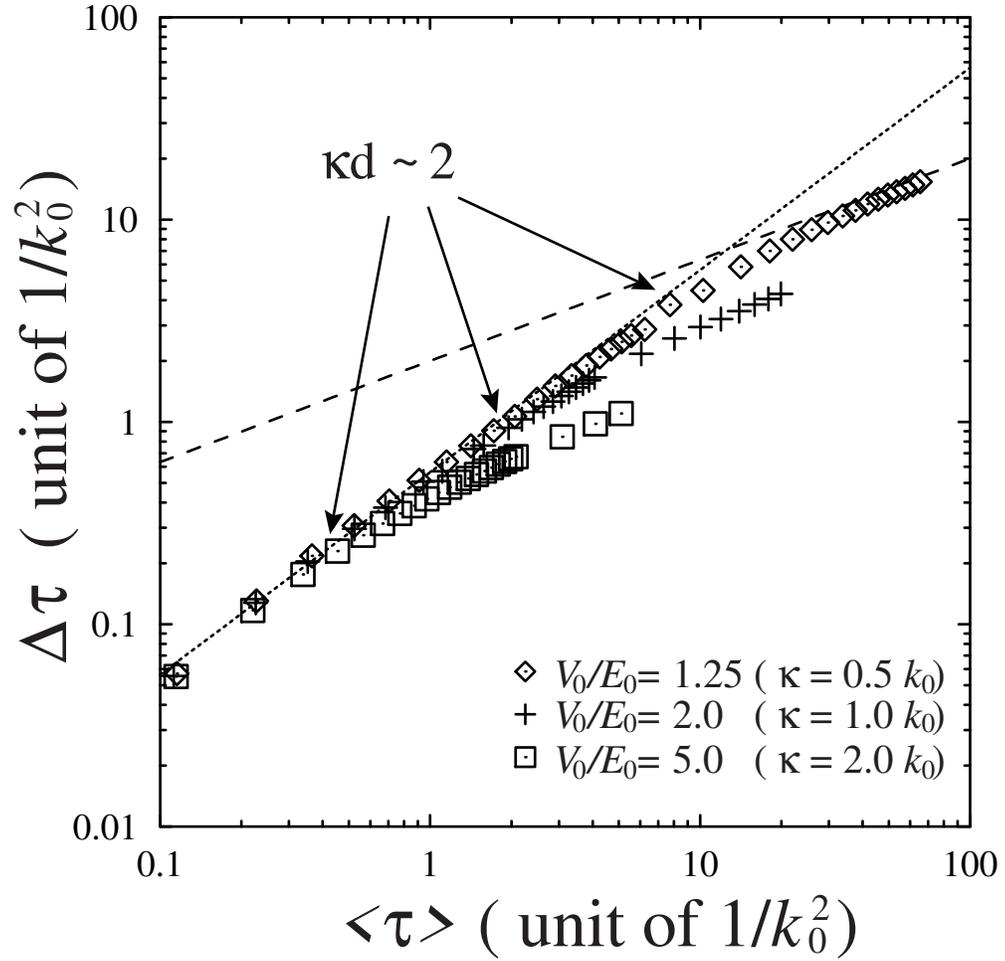}
\caption{\label{fig:6}Relationship between average and deviation 
of tunneling time with the width of wave packet $\Delta x=50/k_0$. 
Dotted line shows the relation $\Delta \tau \propto \langle \tau \rangle$. 
Dashed line shows the relation $\Delta \tau \propto 
\sqrt{\langle \tau \rangle}$.}
\end{figure}

\newpage

\begin{figure}
\includegraphics{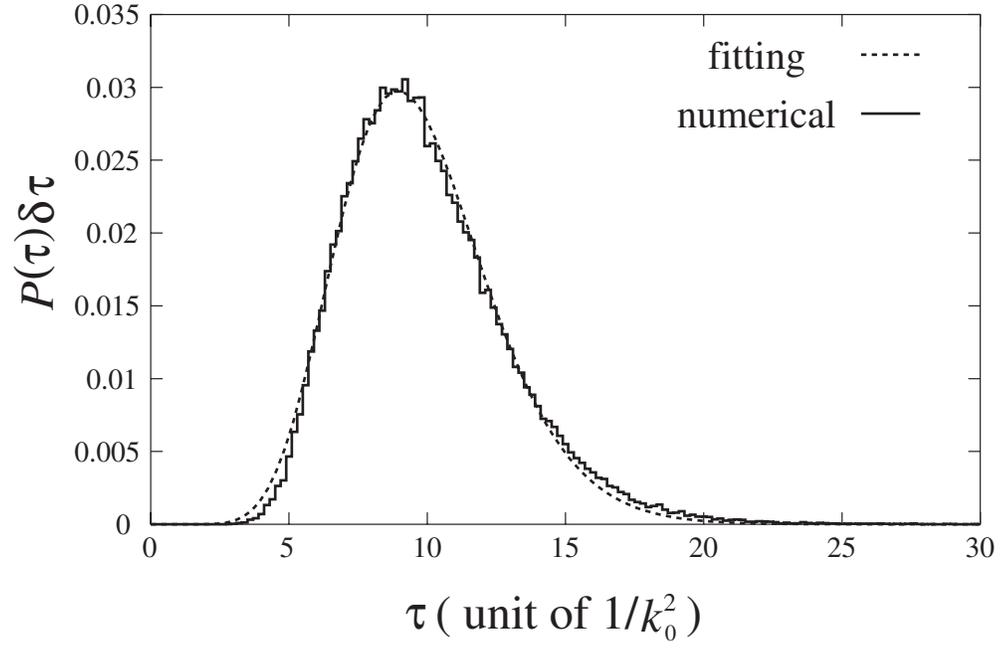}
\caption{\label{fig:7}Example of tunneling time distribution fitted 
by the Gamma distribution in the case of 
$V_0/E_0=2.0, d=10/k_0, \Delta x=50/k_0$.
In this case, we choose the fitting
parameters $ \alpha=11.3 $ and $ \beta = 0.79,$ using least-squares method.}
\end{figure}

\newpage

\begin{figure}
\includegraphics{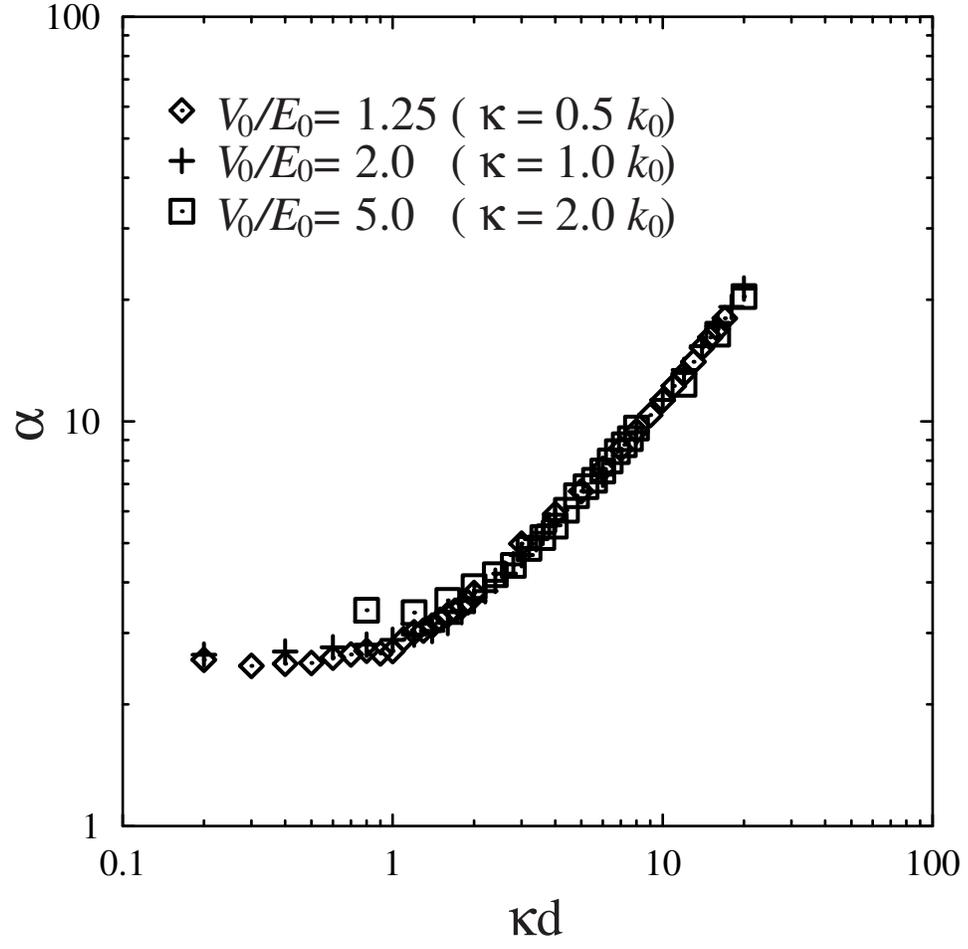}
\caption{\label{fig:8}Fitting parameter $\alpha$ versus 
$\kappa d$ for fixed $\kappa$.}
\end{figure}

\newpage

\begin{figure}
\includegraphics{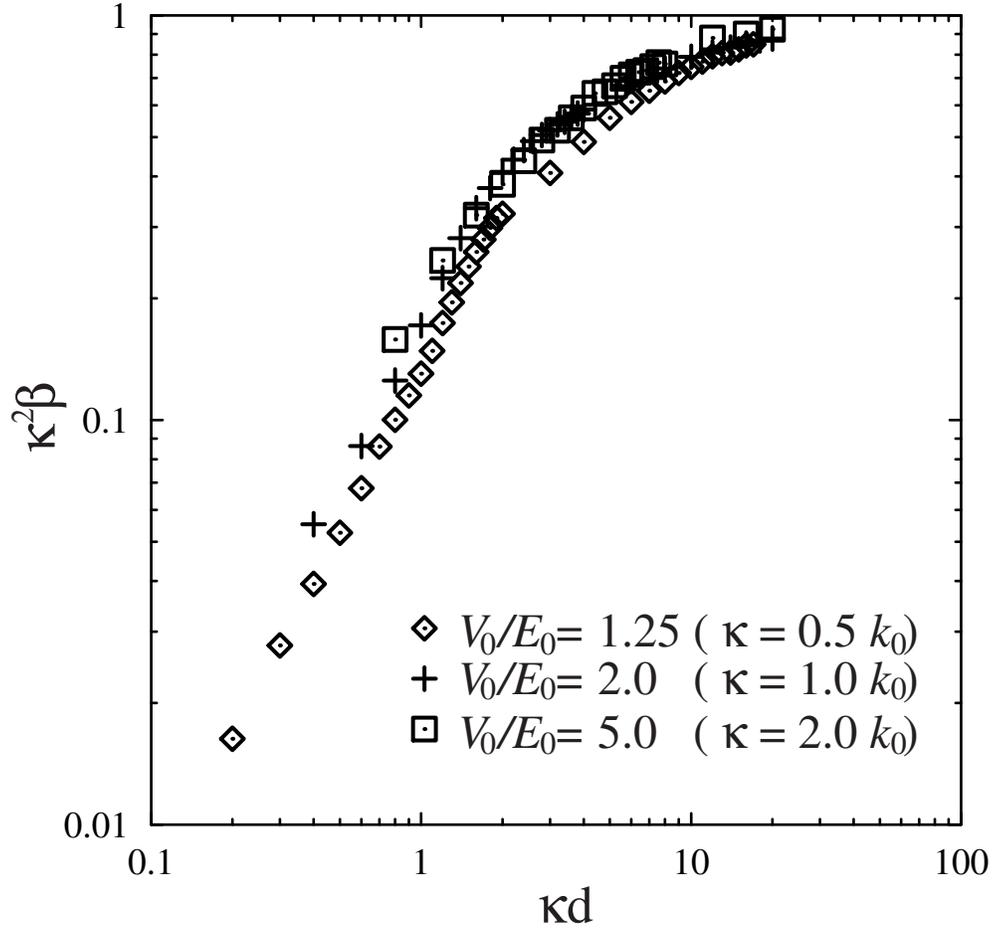}
\caption{\label{fig:9}Fitting parameter $\beta$ multiplied by $\kappa^2$ versus
$\kappa d$ for fixed $\kappa$.}
\end{figure}

\newpage

\begin{figure}
\includegraphics{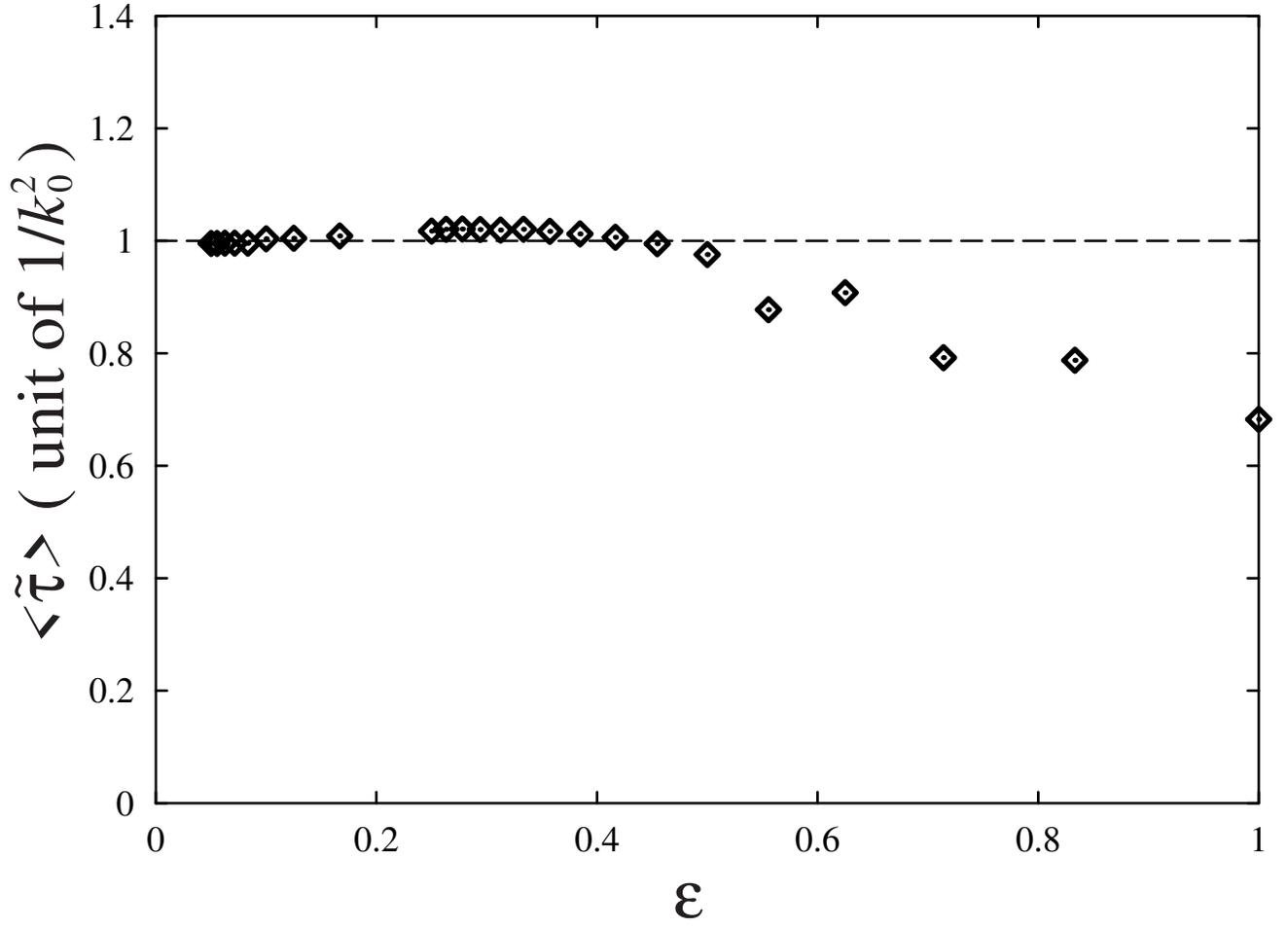}
\caption{\label{fig:10}$\epsilon$-dependence of tunneling time average with potential 
height $V_0/E_0=2.0$ and width $d=1/k_0$ ($\kappa d =1$).
Dashed line shows the WKB time
$\tilde{\tau}_{\rm WKB}=\sqrt{\frac{m}{2(V_0-E_0)}}d$.}
\end{figure}

\newpage

\begin{figure}
\includegraphics{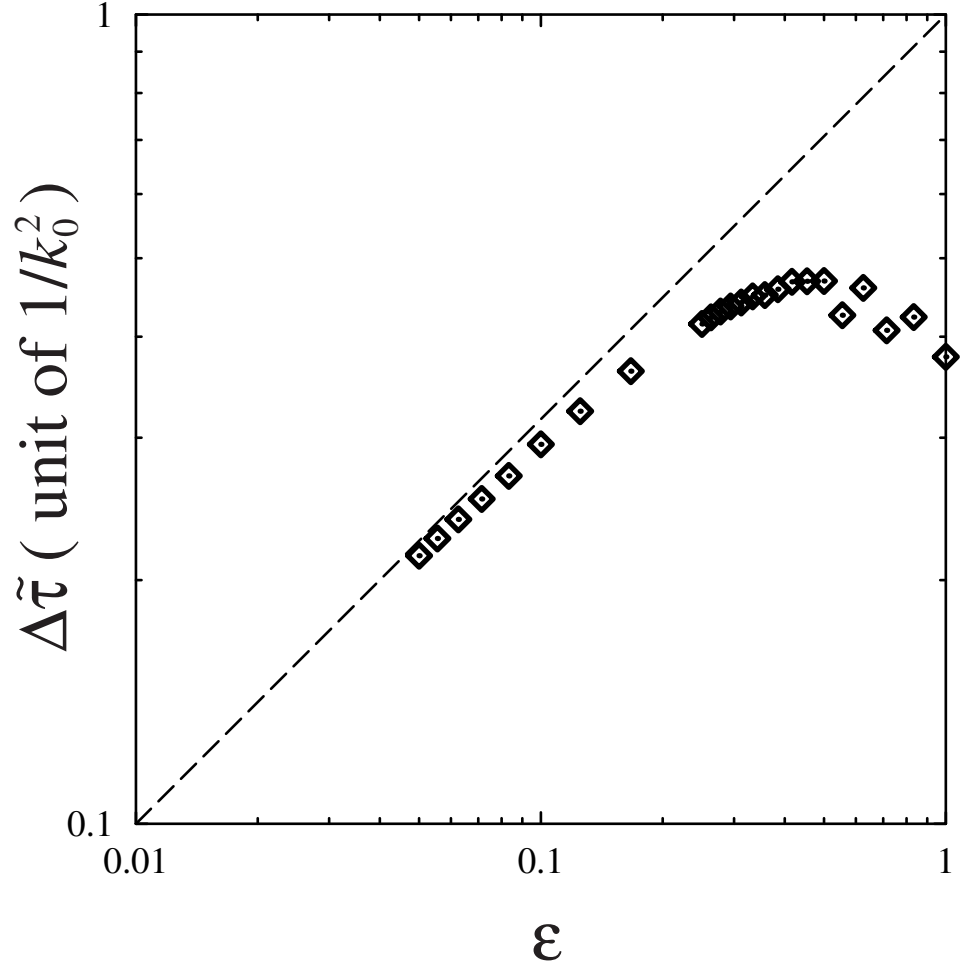}
\caption{\label{fig:11}$\epsilon$-dependence of tunneling time deviation with potential 
height $V_0/E_0=2.0$ and width $d=1/k_0$ ($\kappa d =1$).
Dashed line shows the relation 
$\Delta \tilde{\tau} \propto \sqrt{\epsilon} 
\sqrt{\langle \tilde{\tau} \rangle}$ in the case of particle mode.}
\end{figure}

\end{document}